\documentclass[final]{aipproc}
\layoutstyle{8x11double}
\usepackage{graphicx}
\usepackage{natbib}
\bibpunct{(}{)}{;}{a}{}{,}
\usepackage{graphicx,amsmath}

\begin{document}

\title{Indirect Dark Matter Signals}


\author{Wim de Boer}{
  address={Physikhochhaus, IEKP, Universit\"at Karlsruhe\\
Postfach 6980, D-76128 Karlsruhe\\
Germany\\
e-mail: wim.de.boer@cern.ch}
}
\renewcommand\topfraction{.95}
\renewcommand\bottomfraction{.95}
\renewcommand{\floatpagefraction}{0.9}
\renewcommand{\textfraction}{0.05}

\newcommand{\unity}{\mathbf{1}}
\newcommand{\varunity}{\mbox{\rmfamily 1\hspace{-0.25em}l}}

\newcommand{\Dc}{\mathcal{D}}
\newcommand{\Hc}{\mathcal{H}}
\newcommand{\Lc}{\mathcal{L}}
\newcommand{\Oc}{\mathcal{O}}
\newcommand{\Uc}{\mathcal{U}}

\newcommand{\Rep}{\mbox{Re}}
\newcommand{\Imp}{\mbox{Im}}

\newcommand{\Rm}{\mathbb{R}}

\newcommand{\mfbox}[1]{\fbox{$\displaystyle #1$}}


\newlength{\dslashwidth}
\newcommand{\dslash}[1]{\settowidth{\dslashwidth}{$\diagup$}\mbox{%
\hspace{0.5\dslashwidth}\makebox[0pt]{$#1$}\hspace{-0.5\dslashwidth}%
$\diagup$}}

\newcommand{\bsg}{\ensuremath{b\to X_s\gamma}}
\newcommand{\ch}{\ensuremath{\tilde{\chi}^{\pm}}}
\newcommand{\neu}{\ensuremath{\tilde{\chi}^{0}}}
\newcommand{\sinw}{\ensuremath{\sin^2\theta_W}}
\newcommand{\Mgut}{\ensuremath{M_{\mbox{\scriptsize{GUT}}}}}
\newcommand{\agut}{\ensuremath{\alpha_{\mbox{\scriptsize{GUT}}}}}
\newcommand{\tb}{\ensuremath{\tan\beta}}

\def\aii{\alpha_i^{-1}}
\def\rZ{{\rm Z}}
\def\rW{{\rm W}}
\def\rG{{\rm GUT}}
\def\rS{{\rm SUSY}}
\def\rH{{\rm Higgs}}
\def\rF{{\rm Fam}}
\def\MG{M_\rG}
\newcommand{\mc}{Monte Carlo }
\newcommand{\mcs}{Monte Carlos }
\newcommand{\brem}{brems\-strah\-lung }
\newcommand{\bq}{\begin{equation}}
\newcommand{\eq}{\end{equation}}
\newcommand{\ba}{\begin{array}}
\newcommand{\ea}{\end{array}}
\newcommand{\bqa}{\begin{eqnarray}}
\newcommand{\eqa}{\end{eqnarray}}
\newcommand{\nn}{\nonumber \\}
\newcommand{\mpmm}{\mu^{+}\mu^{-}}
\newcommand{\tptm}{\tau^{+}\tau^{-}}
\newcommand{\sq}{^{2}}
\newcommand{\lnf}{{\ifmmode \Lambda^{(N_f)} \else $\Lambda^{(N_f)}$\fi}}
\newcommand{\ms}{{\ifmmode \overline{MS} \else $\overline{MS}$\fi}}
\newcommand{\dr}{{\ifmmode \overline{DR} \else $\overline{DR}$\fi}}
\newcommand{\lms}{{\ifmmode \Lambda^{(5)}_{\overline{MS}} \else $\Lambda^{(5)}_{\overline{MS}}$\fi}}
\newcommand{\lam}{{\ifmmode \Lambda \else $\Lambda$\fi}}
\newcommand{\gev}{{\ifmmode {\rm GeV} \else ${\rm GeV}$\fi}}
\newcommand{\gevc}{{\ifmmode {\rm GeV/c^2} \else ${\rm GeV/c^2}$\fi}}
\newcommand{\tev}{{\ifmmode {\rm TeV} \else ${\rm TeV}$\fi}}
\newcommand{\tevc}{{\ifmmode {\rm TeV/c^2} \else ${\rm TeV/c^2}$\fi}}
\newcommand{\lp}{{\ifmmode L^+  \else $L^+$\fi}}
\newcommand{\lm}{{\ifmmode L^-  \else $L^-$\fi}}
\newcommand{\mlp}{{\ifmmode M(L^-) \else $M(L^-)$\fi}}
\newcommand{\mlz}{{\ifmmode M(L^0) \else $M(L^0)$\fi}}
\newcommand{\lz}{{\ifmmode L^0 \else $L^0$\fi}}
\newcommand{\ev}{{\ifmmode GeV/c^2 \else $GeV/c^2$\fi}}
\newcommand{\tri}{{\ifmmode \triangleup \else $\triangleup$\fi}}
\newcommand{\unl}{{\ifmmode U_{lL^0} \else $U_{lL^0}$\fi}}\newcommand{\gL}{{\ifmmode g_L \else $g_{L}$\fi}}
\newcommand{\gR}{{\ifmmode g_R  \else $g_{R}$\fi}}
\newcommand{\gumu}{{\ifmmode \gamma^{\mu} \else $\gamma^{\mu}$\fi}}
\newcommand{\gunu}{{\ifmmode \gamma^{\nu} \else $\gamma^{\nu}$\fi}}
\newcommand{\gdmu}{{\ifmmode \gamma_{\mu} \else $\gamma_{\mu}$\fi}}
\newcommand{\gdnu}{{\ifmmode \gamma_{\nu} \else $\gamma_{\nu}$\fi}}
\newcommand{\stw}{{\ifmmode\sin^2\theta_W \else $\sin^{2}\theta_{W}$ \fi}}
\newcommand{\sws}{{\ifmmode \;\sin^2\theta_W  \else $\;\sin^{2}\theta_{W}$ \fi}}
\newcommand{\cws}{{\ifmmode \;\cos^2\theta_W  \else $\;\cos^{2}\theta_{W}$ \fi}}
\newcommand{\sw}{{\ifmmode \;\sin\theta_W  \else $\sin\theta_{W}$ \fi}}
\newcommand{\cw}{{\ifmmode \;\cos\theta_W  \else $\;\cos\theta_{W}$ \fi}}
\newcommand{\tw}{{\ifmmode \;\tan\theta_W  \else $\;\tan\theta_{W}$ \fi}}
\newcommand{\qq}{{\ifmmode q\overline{q} \else $q\overline{q}$\fi}}
\newcommand{\lR}{{\ifmmode l_R \else $l_R$\fi}}
\newcommand{\lL}{{\ifmmode l_L \else $l_L$\fi}}
\newcommand{\nt}{{\ifmmode \nu_{\tau} \else $\nu_{\tau}$\fi}}
\newcommand{\nuR}{{\ifmmode \nu_R  \else $\nu_R$\fi}}
\newcommand{\nuL}{{\ifmmode \nu_L  \else $\nu_L$\fi}}
\newcommand{\qR}{{\ifmmode g_R  \else $q_R$\fi}}
\newcommand{\qL}{{\ifmmode q_L  \else $q_L$\fi}}
\newcommand{\qRp}{{\ifmmode q_R'  \else $q_{R}$'\fi}}
\newcommand{\qLp}{{\ifmmode q_L'  \else $q_{L}$'\fi}}
\newcommand{\est}{{\ifmmode e^{\bf \ast} \else $e^{\bf \ast}$\fi}}
\newcommand{\lst}{{\ifmmode l^{\bf \ast} \else $l^{\bf \ast}$\fi}}
\newcommand{\must}{{\ifmmode \mu^{\bf \ast} \else $\mu^{\bf \ast}$\fi}}
\newcommand{\taust}{{\ifmmode \tau^{\bf \ast} \else $\tau^{\bf \ast}$ \fi}}
\newcommand{\pperp}{{\ifmmode p_t  \else $p_t$\fi}}
\newcommand{\et}{{\ifmmode E_t  \else $E_t$\fi}}
\newcommand{\xt}{{\ifmmode x_t  \else $x_t$\fi}}
\newcommand{\smumu}{{\ifmmode \sigma_{\mu\mu}  \else $\sigma_{\mu\mu}$ \fi}}
\newcommand{\eg}{{\ifmmode e\gamma  \else $e\gamma$\fi}}
\newcommand{\epem}{{\ifmmode e^+e^-  \else $e^+e^-$\fi}}
\newcommand{\lplm}{{\ifmmode L^+L^-  \else $L^+L^-$\fi}}
\newcommand{\pp}{{\ifmmode p\overline p  \else $p\overline p$\fi}}
\newcommand{\llz}{{\ifmmode L^0\overline{L}^0 \else $L^0\overline{L}^0$\fi}}
\newcommand{\epemt}{{\ifmmode e^+e^- \to  \else $e^+e^- \to$\fi}}
\newcommand{\eb}{{\ifmmode E_{beam}  \else $E_{beam}$\fi}}
\newcommand{\ip}{{\ifmmode pb^{-1}  \else $pb^{-1}$\fi}}
\newcommand{\upm}{{\ifmmode ^{\pm}  \else $^{\pm}$\fi}}
\newcommand{\de}{{\ifmmode ^{\circ}  \else $^{\circ}$ \fi}}
\newcommand{\appr}{{\ifmmode \sim \else $\sim$ \fi}}
\newcommand{\corresp}{{\ifmmode \stackrel{\wedge}{=} \else $\stackrel{\wedge}{=}$ \fi}}
\newcommand{\sqrts}{{\ifmmode \sqrt{s} \else $\sqrt{s}$\fi}}
\newcommand{\zz}{{\ifmmode Z^0  \else $Z^0$\fi}}
\newcommand{\mz}{{\ifmmode M_{Z}  \else $M_{Z}$\fi}}
\newcommand{\mzs}{{\ifmmode M_{Z}^2  \else $M_{Z}^2$\fi}}
\newcommand{\mw}{{\ifmmode M_{W}  \else $M_{W}$\fi}}
\newcommand{\mws}{{\ifmmode M_{W}^2  \else $M_{W}^2$\fi}}
\newcommand{\mh}{{\ifmmode M_{Higgs}  \else $M_{Higgs}$\fi}}
\newcommand{\gt}{{\ifmmode \Gamma_{tot} \else $\Gamma_{tot}$\fi}}
\newcommand{\msusy}{{\ifmmode M_{SUSY}  \else $M_{SUSY}$\fi}}
\newcommand{\msusys}{{\ifmmode M_{SUSY}^2  \else $M_{SUSY}^2$\fi}}
\newcommand{\su}{{\ifmmode SU(3)_C\otimes\- SU(2)_L\otimes\- U(1)_Y \else $SU(3)_C\otimes SU(2)_L\otimes U(1)_Y$\fi}}
\newcommand{\suthree}{{\ifmmode SU(3)_C  \else $SU(3)_C$\fi}}
\newcommand{\sutwo}{{\ifmmode  SU(2)_L\otimes U(1)_Y \else $SU(2)_L\otimes U(1)_Y$\fi}}
\newcommand{\taup} {{\ifmmode \tau_{proton} \else $\tau_{proton}$\fi}}
\newcommand{\as}{{\ifmmode \alpha_{s}  \else $\alpha_{s}$\fi}}
\newcommand{\mgut}{{\ifmmode M_{GUT}  \else $M_{GUT}$\fi}}
\newcommand{\mguts}{{\ifmmode M_{GUT}^2  \else $M_{GUT}^2$\fi}}
\newcommand{\mze} {{\ifmmode m_0        \else $m_0$\fi}}
\newcommand{\mha}{{\ifmmode m_{1/2}    \else $m_{1/2}$\fi}}
\newcommand{\mb} {{\ifmmode m_{b}    \else $m_{b}$\fi}}
\newcommand{\mt} {{\ifmmode m_{t}    \else $m_{t}$\fi}}
\newcommand{\mts} {{\ifmmode m_{t}^2    \else $m_{t}^2$\fi}}
\newcommand {\rb}[1]{\raisebox{1.5ex}[-1.5ex]{#1}}
\newcommand{\mtau}{{\ifmmode m_{\tau}  \else $m_{\tau}$\fi}}
\newcommand{\dpp}{{\ifmmode \delta_{pert} \else $\delta_{pert}$\fi}}
\newcommand{\dnp}{{\ifmmode\delta_{non-pert}\else$\delta_{non-pert}$\fi}}
\newcommand{\dew}{{\ifmmode \delta_{\rm EW}\else $\delta_{\rm EW}$\fi}}
\newcommand{\rt}{{\ifmmode R_{\tau}  \else $R_{\tau} $\fi}}
\newcommand{\rz}{{\ifmmode R_{Z}  \else $R_{Z} $\fi}}
\newcommand{\into}{\rightarrow}
\newcommand{\SM}{Standard Model}
\newcommand{\swb}{{\ifmmode \sin^2\theta_{\overline{MS}} \else $\sin^2\theta_{\overline{MS}}$\fi}}
\newcommand{\cwb}{{\ifmmode \cos^2\theta_{\overline{MS}} \else $\cos^2\theta_{\overline{MS}}$\fi}}
\newcommand{\ttbs}{\char'134}
\newcommand{\besg}{$b  \to  X_s \gamma~ $}
\newcommand{\mzero}{\rm m_0}
\newcommand{\mhalf}{\rm m_{1/2}}

\begin{abstract}

Dark Matter annihilation (DMA)
may yield an excess  of
gamma rays and antimatter particles, like antiprotons and positrons,
above the background from cosmic ray interactions.
Several signatures, ranging from the positron excess, as observed by HEAT, AMS-01 and PAMELA,
the gamma ray excess, as observed by the  EGRET spectrometer, the WMAP-haze, and constraints from
antiprotons, as observed by CAPRICE, BESS and PAMELA, have been discussed in the literature.
Unfortunately, the different signatures all lead to different WIMP masses, indicating that at least some of these interpretations are likely to be incorrect. Here we review them and discuss their relative merits
and uncertainties. New x-ray data from ROSAT suggests non-negligible convection in our Galaxy, which leads
to an order of magnitude uncertainty in the yield of charged particles from DMA, since even a rather small convection will let drift the charged particles in the halo to outer space.

\end{abstract}

\maketitle

\section{Introduction}

In this plenary talk contribution to SUSY08 we update the status of the possible Dark Matter annihilation (DMA) signatures, as they have been discussed at the last SUSY07 conference in Karlsruhe \cite{hoopersusy07}. These include the EGRET excess of gamma rays, both in the Galactic \cite{us} and extragalactic component \cite{elsaesser,deboer_ex}, the  increase in the high energy positron fraction, as observed by HEAT \cite{heat}, AMS-01 \cite{ams01} and preliminary by PAMELA \cite{pamela}, the positron excess towards the Galactic center by the INTEGRAL satellite \cite{integral,integral_nature}, the excess in synchrotron radiation towards the Galactic center, known as the WMAP-haze \cite{haze,haze1} and the constraints from antiprotons, as observed by the balloon experiments CAPRICE \cite{caprice} and  BESS \cite{bess} and the space experiment PAMELA\cite{pamela}. 

Before starting these discussions it is worthwhile to mention the results of two recent papers, which are of fundamental importance for indirect detection. The first one concerns the results from the X-ray spectra in our Galaxy, as measured with the ROSAT satellite \cite{rosat}, which provides evidence for convective winds in the halo with speeds of the order of 150-400 km/s, as discussed also in a recent Nature article \cite{breitschwerdt_nature}. Note that these are speeds at the lower edges of convective winds:   in star-burst driven galaxies winds  as high as 3000 km/s have been observed.
 However, even such a rather modest convection is enough to let the charged particles in the halo be captured in the plasma and drift towards outer space, thus reducing the DMA contribution of charged particles drastically. This will be discussed in detail later.

The second paper concerns the halo profile, which is estimated from N-body simulations to have a steep cuspy profile towards the Galactic center, as was first discussed by Navarro, Frenk and White and therefore usually referred to as the NFW-profile \cite{nfw}. Higher resolution simulations show that the clumps in the dark matter (DM) distribution, which boost the DMA by the so-called boost factor because of the local higher density, are distributed more like a cored distribution of the Einasto shape \cite{springel}. Therefore the DMA signal is {\it not} expected to have a strong peak towards the center, as assumed in many  studies, see some reviews \cite{reviews}.


First the excess of energetic gamma rays will be discussed, since these are not influenced by magnetic fields. Then the possible scenarios for propagation models are discussed including the influence from convection consistent with ROSAT data \cite{rosat}.
Given the large uncertainties from the propagation models it is shown that all signatures from DMA based on charged particles have large background uncertainties from the presence of convection.

\begin{figure}
{\includegraphics[width=0.346\textwidth,height=0.3\textwidth,clip]{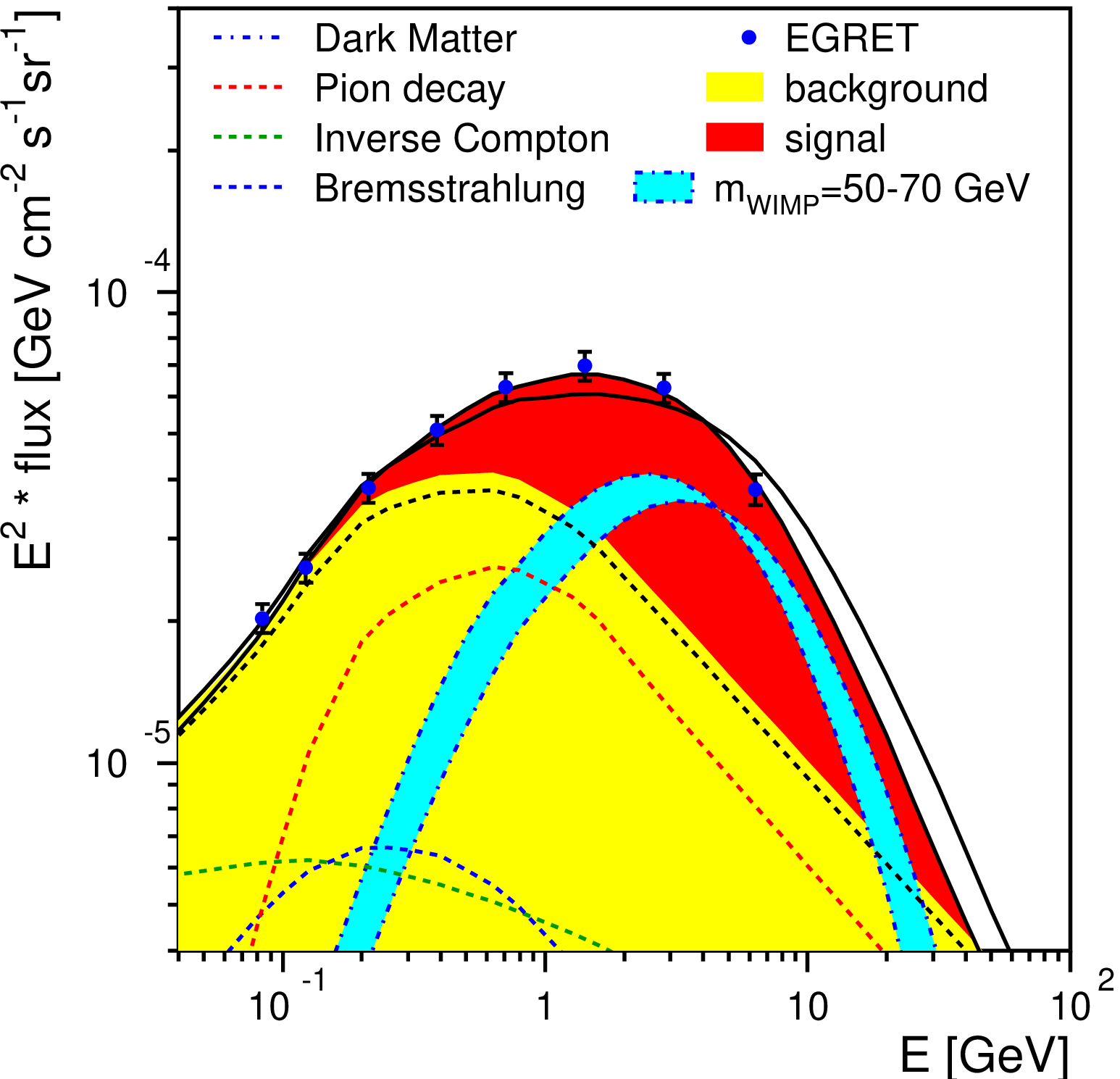}}
{\includegraphics[width=0.346\textwidth,height=0.3\textwidth,clip]{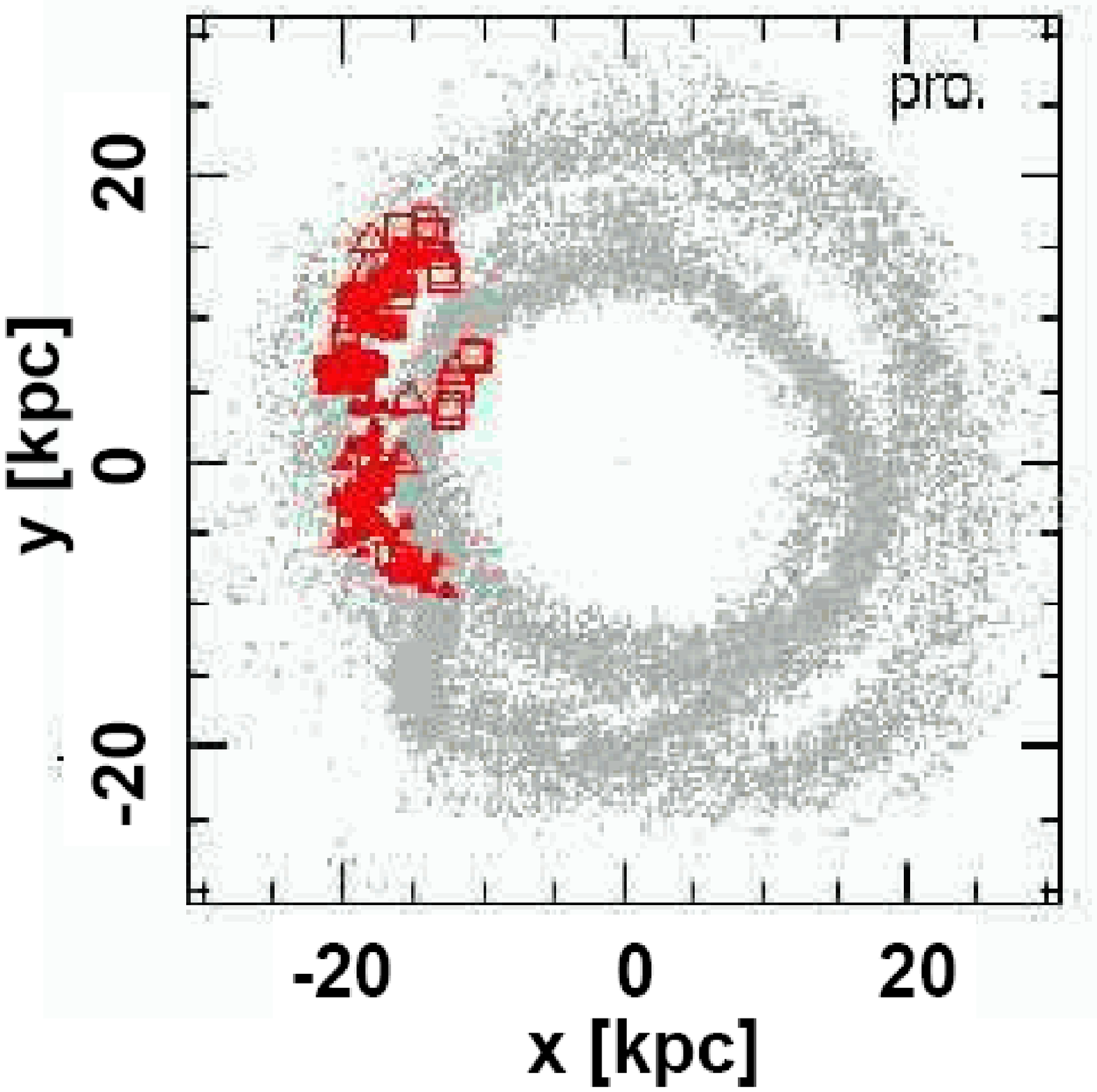}}
{\includegraphics[width=0.346\textwidth,height=0.33\textwidth,clip]{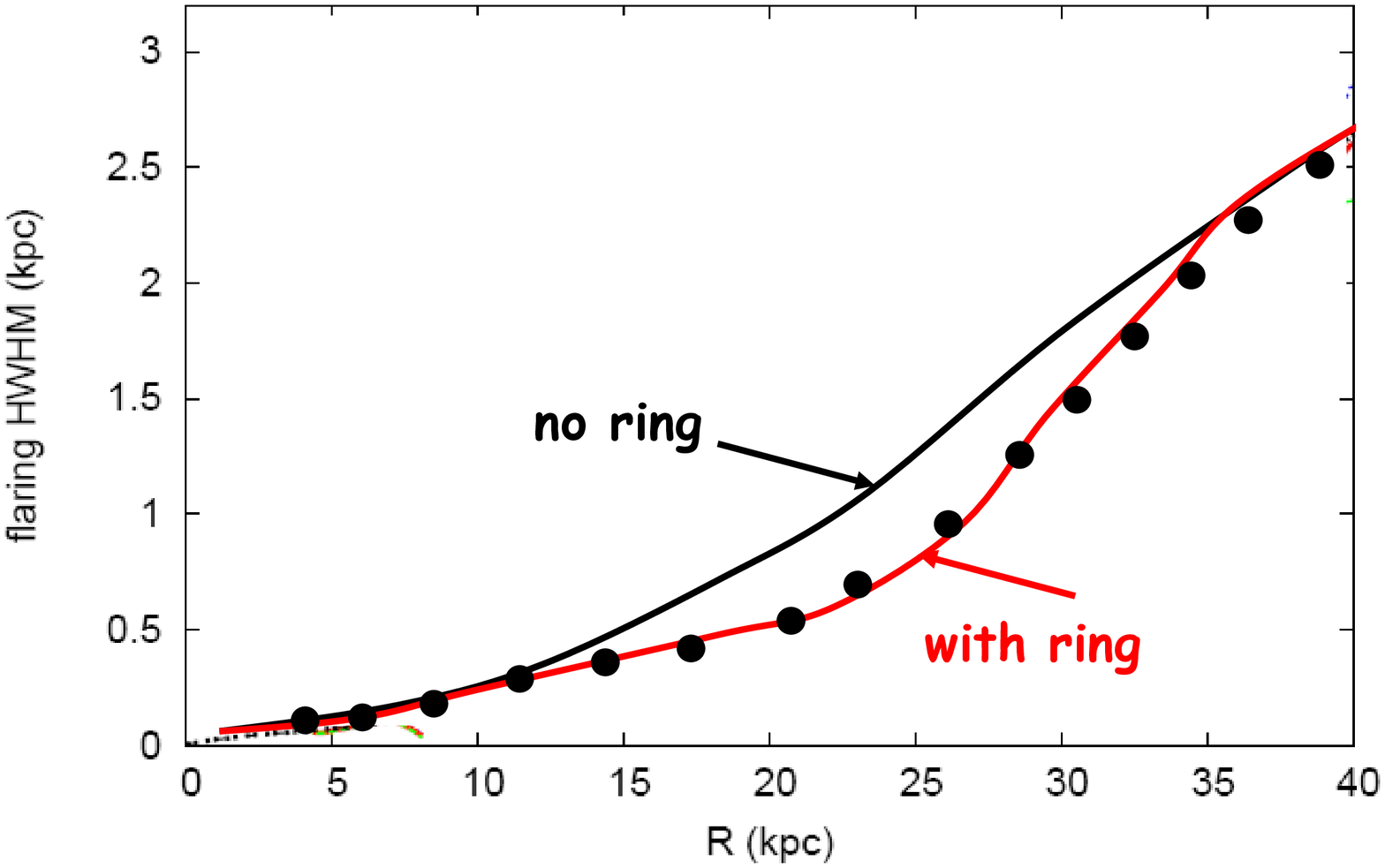}}
\caption{Left: Fit of the shapes of background and DMA signal to the EGRET data in the direction of the Galactic center. The light shaded (yellow) area indicates the background using the shapes known from accelerator experiments, while the dark shaded (red) area corresponds to the signal contribution from DMA for a 60 GeV WIMP mass, where the small intermediate (blue) shaded area corresponds to a variation of the WIMP mass between 50 and 70 GeV. Centre:  Results of an N-body simulation of the tidal disruption of the
Canis Major dwarf Galaxy, whose orbit was fitted to the observed stars (red points). The simulation predicts a ringlike structure of dark matter with a radius of 13 kpc. From \cite{penarrubia}.
Right: The half-width-half-maximum (HWHM) of the gas layer of the Galactic disk as function of the distance from the Galactic center. Clearly, the fit including a ring of dark matter above 10 kpc describes the data much better. Adapted from data in \cite{kalberla}.
\label{f1}}
\end{figure}
\section{The EGRET excess of diffuse Galactic gamma rays}\label{egret}

An excess of diffuse gamma rays  has
 been observed by the EGRET telescope on board of NASA's CGRO (Compton Gamma
Ray Observatory)\cite{hunter}. Below 1 GeV the cosmic ray (CR)
interactions describe the data  well, but above 1 GeV the data are up to a
factor two above the expected background. The excess shows all the features of DMA
 for a WIMP mass between 50 and 70 GeV, as shown in Fig. \ref{f1} \cite{us}.
These features include fulfillment of the two
basic, but very constraining conditions  expected from any indirect DMA signal: (i) the
excess should have the same {\it spectral shape} in all sky directions. (ii) the
excess should be observable in a large fraction of the sky with an {\it intensity
distribution} corresponding to the gravitational potential of our Galaxy. The
latter means that one should be able to relate the distribution of the excess to
the rotation curve. Both conditions are indeed met by
the EGRET data \cite{us}. In addition, the results are perfectly consistent with
the expectations from Supersymmetry\cite{egret_susy}.

The  analysis of the EGRET data was performed with a so-called data-driven calibration of the background,
a procedure commonly used in accelerator experiments to reduce the sensitivity to model dependence of signal and background calculations. Such analysis techniques are rather unusual in the astrophysics community or among theorists,
who typically use the standard procedure of taking a Galactic model to calculate the background and a
 certain DM halo model to calculate the signal and then compare signal plus background with data. Usually an NFW DM profile is used, which gives the highest DMA signal rate, but should be
abandoned according to the newer high resolution N-body simulations mentioned in the introduction.
Such analysis are highly sensitive to uncertainties in the background models.
\begin{figure}
\includegraphics[width=0.45\textwidth,height=0.4\textwidth,angle=0]{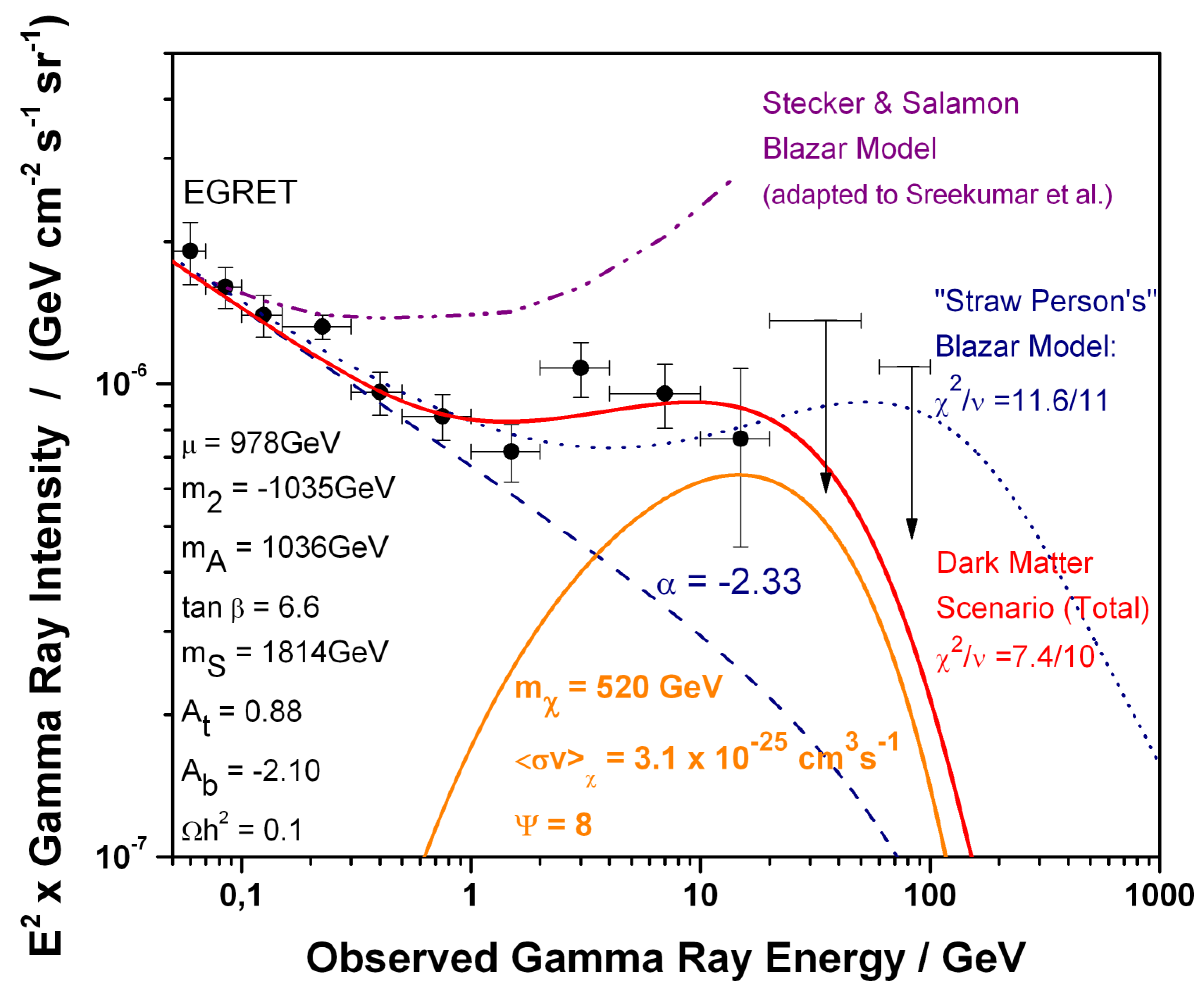}\\
\includegraphics[width=0.45\textwidth,height=0.4\textwidth,angle=0]{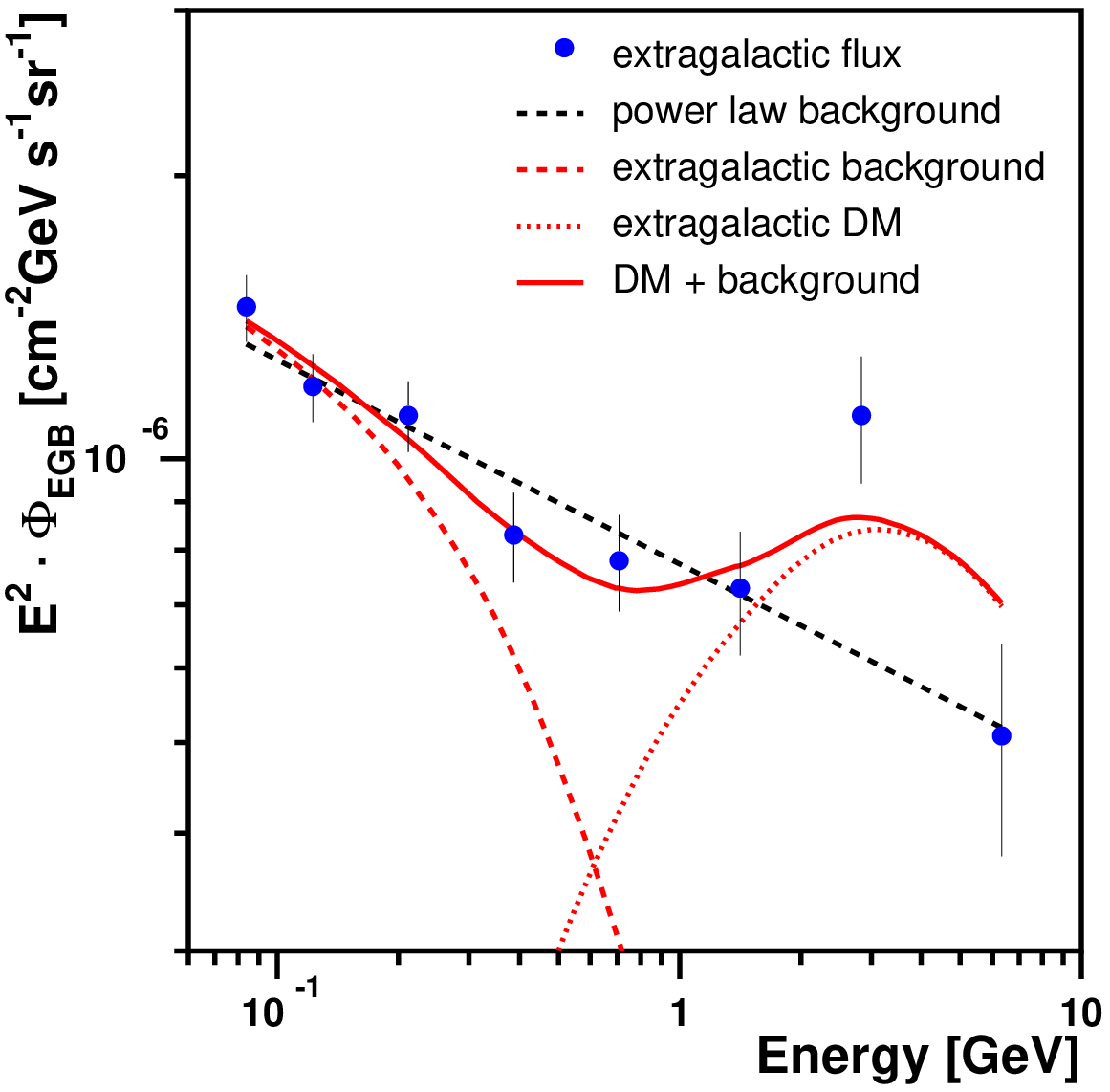}
\caption{The extragalactic gamma ray spectrum assuming a Galactic foreground without DMA (left) and with (right) DMA. From Refs. \cite{elsaesser} and \cite{deboer_ex}, respectively. The long tail observed, if no DMA in the Galactic foreground is assumed, needs a WIMP mass of 520 GeV, while including  DMA in the foreground results in a  WIMP mass compatible with the Galactic EGRET excess, as can be seen from a comparison with the left panel of  Fig. \ref{f1}.}
\label{f2}
\end{figure}

A data-driven approach is particularly suitable for the analysis of gamma rays, since the shape of the dominant background, which is the $\pi^0$ production in inelastic collisions of CR protons on the hydrogen gas of the disk, is well known from so-called fixed target accelerator experiments in which a proton beam is scattered on a hydrogen target \cite{pdb}.
Furthermore, the shape of the DMA signal is known from $e+e-$ annihilation, so the gamma ray shapes of both, signal and background are known from accelerators with high precision, since these reactions happen to be the best studied ones in high energy physics \cite{pdb}. Since the signal has a significantly harder spectrum than the background one can perform a data-driven analysis by simply fitting the two shapes to the experimental data
with a free normalization for each shape,  thus obtaining the absolute contribution from signal and background for each sky direction in a rather model-independent way. Uncertainties in the interstellar background shape arise from solar modulation and in addition from the uncertainties from electron CRs generating gamma-rays by inverse Compton scattering and Bremsstrahlung. However, since the electron flux of CRs is two orders of magnitude below the proton flux, this effect is small
in the region of interest above 1 GeV. If one allows the interstellar CR spectral shape to be different from the locally observed shape a larger uncertainty is obtained and the EGRET excess can be reduced signi\-ficantly \cite{us}.
However, two points should be noted: i) the EGRET data  are well reproduced {\it in all sky directions} by the background shape using the spectral shape of the {\it local} spectra, if DMA is included; ii) the spatial distribution in the sky is not effected by a different spectral shape, even if the absolute value of the excess is, since the overall distribution is normalized to the mass needed for the rotation speed of the solar system, so a different excess is renormalized to the total mass.

 A data driven method  reduces the sensitivity to calibration errors, which have been proposed in Ref. \cite{stecker} as an explanation for the EGRET excess. The calibration was modified in such a way to bring the excess to zero, implying  a much larger correction than expected from the quoted errors. 
 However,
the spatial distribution of the excess and the correlation with the rotation curve and gas flaring, which form the hallmark of the DMA interpretation, as will be discussed below, are
largely independent of  calibration errors as long as the calibration is similar in all sky directions.
Although there is
some uncertainty in the efficiency of the EGRET veto counter at higher energies because
of the backsplash from the calorimeter, this effect should not start at 1 GeV and is not seen
in recent more detailed simulations \cite{egret_baughman}.

The average $\chi^2$ per degree of freedom summed over all ca. 1400 data points is around 1, indicating that the errors are correctly estimated. But above all such a good $\chi^2$ implies that the main conditions for a signal of DMA are fulfilled, namely i) the {\it shape} of the excess should correspond to the fragmentation of mono-energetic quarks with the same energy in {\it all} sky directions and ii) the {\it intensity} distribution of the excess  agrees with the rotation curve.

The results on the spatial distribution are surprising: the background agrees within errors with the expectation from  GALPROP, the most up-to-date Galactic propagation model \cite{galprop}, as can be seen from Fig. 3 in Ref. \cite{us}, but the derived halo profile  shows some unexpected substructure: outside the disk it corresponds to a cored halo profile, as expected from the new N-body simulations, but inside the disk it reveals two additional doughnut-like structures at distances of about 4 and 13 kpc from the Galactic center.  Ringlike structures are expected from the tidal disruption of dwarf galaxies captured in the gravitational field of our Galaxy. 
The "ghostly" ring of stars or Monocerus stream (with about $10^8-10^9$ solar masses in visible matter) could be the tidal streams of the Canis Major dwarf galaxy (see e.g. \cite{cma,penarrubia} and references therein). If so, the tidal streams  predicted from N-body simulations are perfectly consistent with the ring at 13 kpc  \cite{penarrubia}, as shown in the central panel of Fig. \ref{f1}.  The strong gravitational potential well in this stream was  confirmed from the gas flaring, which is reduced at the position of the ring \cite{kalberla}.
The half-width-half-maximum of the gas layer in the disk is shown on the right-hand panel of Fig. \ref{f1}. The reduced gas flaring corresponds to more than $10^{10}$ solar masses, in  agreement with the EGRET ring. It should be noted that the peculiar shape of the gas flaring was only understood after the astronomers heard about the EGRET ring. The effect is so large that visible matter cannot explain this peculiar shape. Also the peculiar change in slope of the rotation curve can only be explained by a ringlike structure \cite{us}. A similar ring
in the outer disk has been discovered in a nearby galaxy, indicating that such infalls may shape
the disk and its warps \cite{penarrubia1}.

So the DMA interpretation of the EGRET excess at 13 kpc is strongly supported by these independent astronomical observations. The ring at 4 kpc might also originate from the disruption of a smaller dwarf galaxy, but here the density of stars is too high to find evidence for tidal streams. However, direct evidence of a stronger gravitational potential well in this region comes from the ring of dust at this location. Since this ring is slightly tilted with respect to the plane its presence and orientation are most easily explained by the presence of a ringlike structure of DM.
It should be noted that such structures can only by discovered by a  model-independent data-driven approach. 
\section{Extragalactic Gamma Rays}

A small fraction of the diffuse gamma rays stems from outside the Galaxy. This extragalactic background is difficult to determine, since it requires a subtraction of the Galactic foreground from the data. Since the EGRET excess is not well described by GALPROP, the extragalactic background in this region shows a tail, which has been discussed as a DMA signal \cite{elsaesser}. However, if the DMA is included in the foreground, the tail largely disappears and the data can be fitted with the sum of a shape typical of point sources plus a DMA contribution for a WIMP mass
consistent with the Galactic DMA signal \cite{deboer_ex}.
The distributions of the extragalactic gamma rays in both analysis are compared in Fig. \ref{f2}.

\section{Propagation models including convection}
\begin{figure}
\includegraphics[width=0.35\textwidth,height=0.35\textwidth,clip]{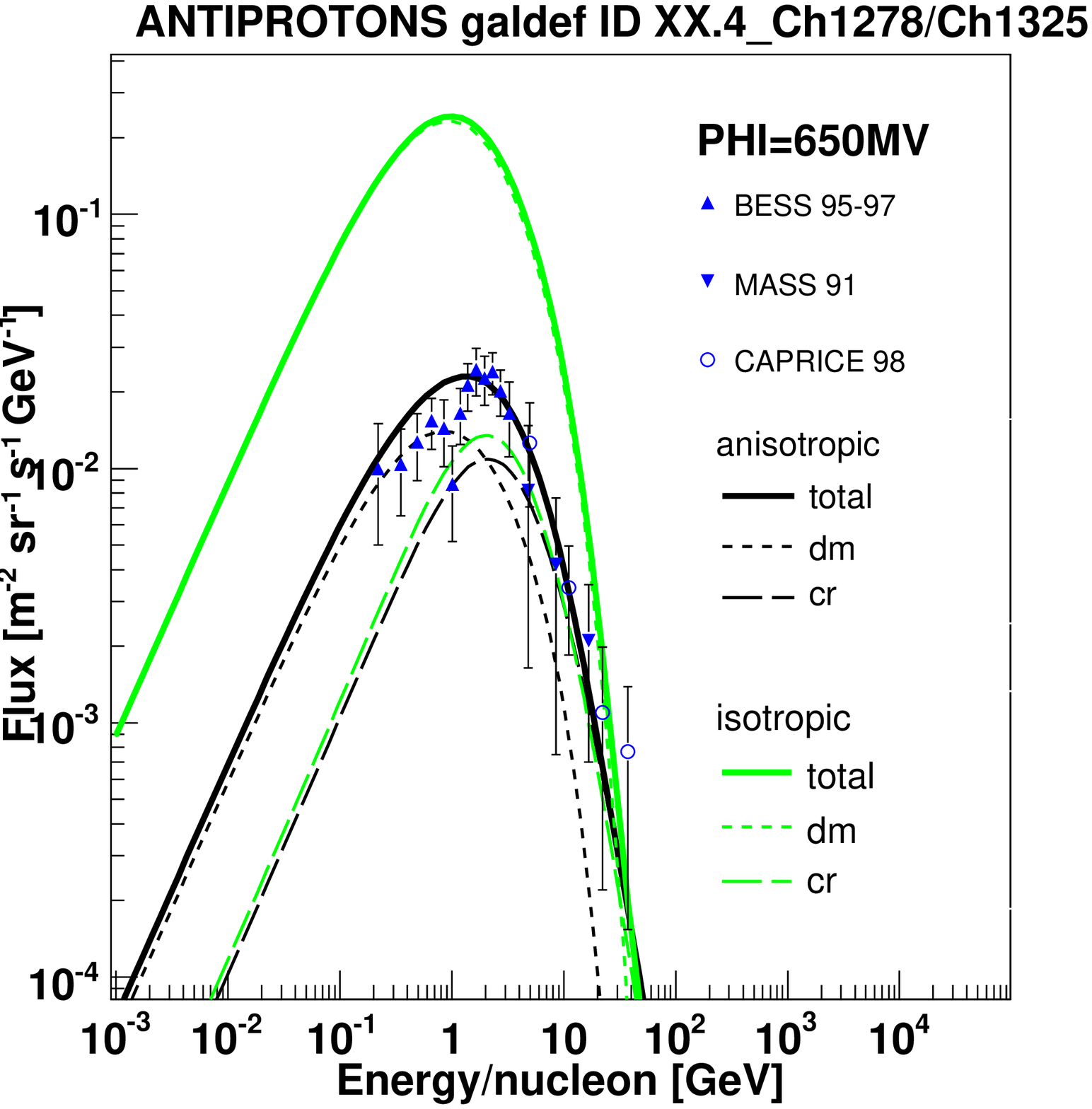}
\includegraphics[width=0.35\textwidth,height=0.35\textwidth,clip]{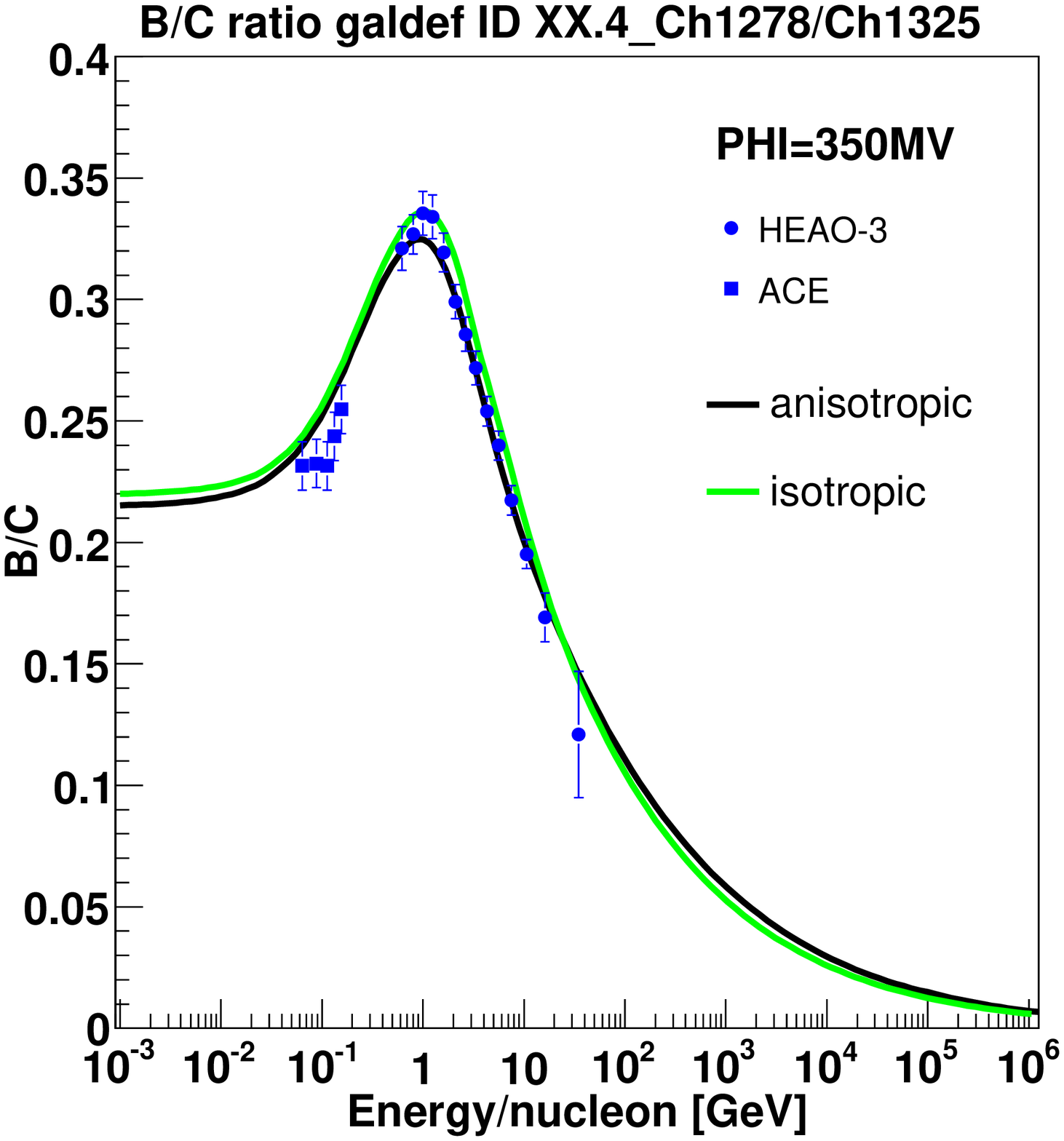}
\includegraphics[width=0.35\textwidth,height=0.35\textwidth,clip]{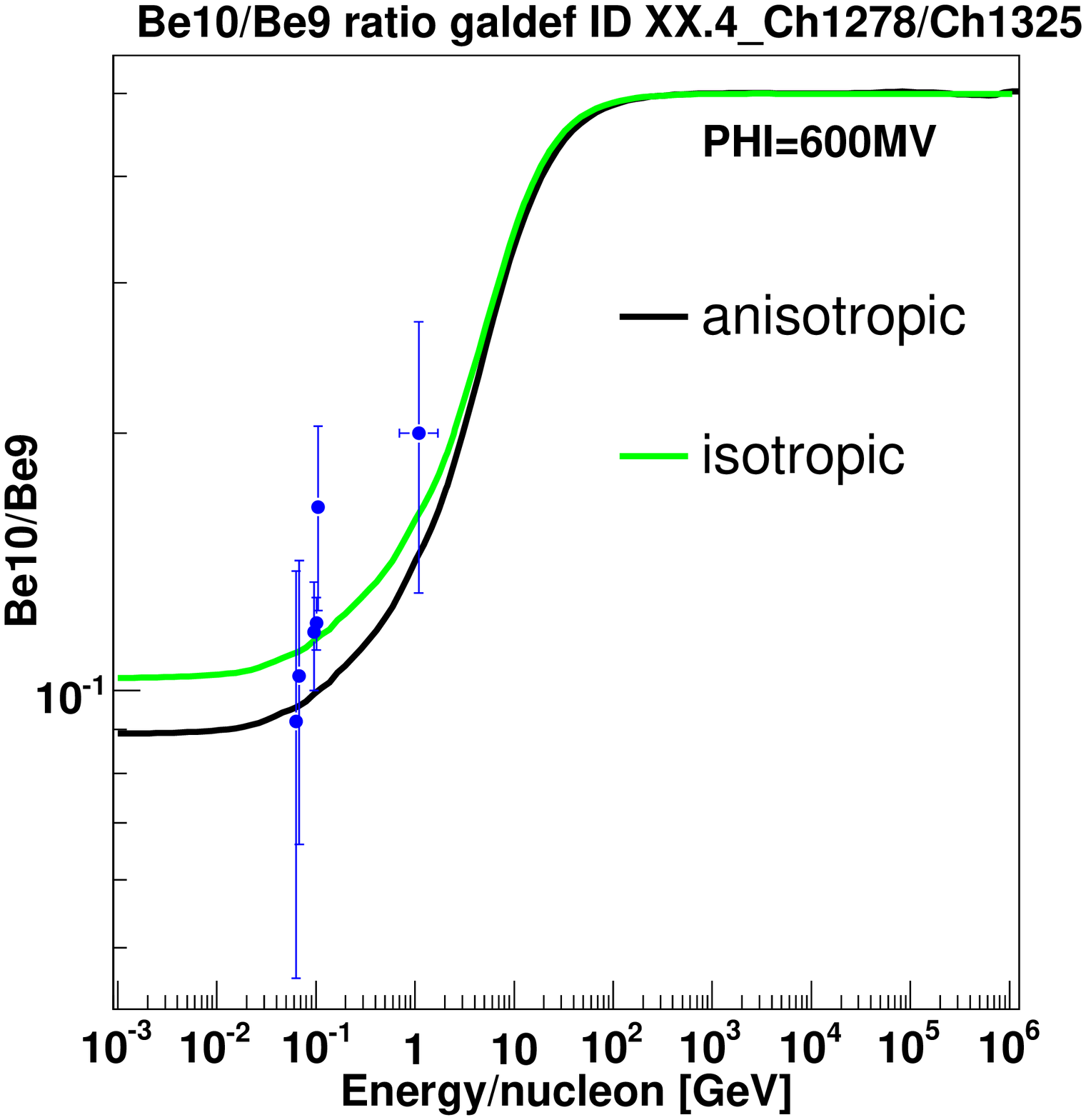}
\caption{Results from the anisotropic and isotropic propagation models (including DMA) for
antiprotons, $B/C$ and $^{10}Be/^9Be$.}
\label{f3}
\end{figure}

 High energy cosmic rays (CRs) travel close to the speed of light. Nevertheless,
their averaged speed is much reduced by the interaction with the plasma created by the
ensemble of CRs in the Galaxy. The resulting propagation is usually described as a diffusion process
combined with a transport perpendicular to the disk by Galactic winds originating from
Supernovae (SN) explosions. The SN explosions are  thought to accelerate charged particles to become CRs.
The scattering on the magnetic turbulence of the plasma tends to zero, if the energy E of the particles
approaches zero, so the diffusion coefficient is proportional to $\beta E^\alpha$ with $0.3 \le \alpha \le 0.6$ ($\beta =v/c$)
according to theoretical expectations and experimental observations.  The convection by Galactic winds
is energy independent and for  observed convection speeds above 150 km/s it is significant for particles below the TeV range, at least in the regions where SN explosions
are frequent. Details about isotropic propagation models can be found in a recent review \cite{strong_rev}.
Details on anisotropic propagation models will be published elsewhere \cite{gebauer}.

Isotropic propagation models cannot include convection speeds above ca. 10 km/s, since in these models secondary particles, like Boron, are produced by the fragmentation of heavier nuclei (mainly C, N, O) hitting the gas of the target. Therefore too high convection lets drift the CRs away from the disk, thus producing too few secondaries.  A solution would be to allow   diffusion in the disk to be much slower than in the halo, thus increasing the CR density in the disk. There are several reasons why this could be so. E.g. it may be that particles diffuse predominantly along the regular magnetic fields of the spiral arms, thus having a reduced diffusion speed in the radial direction. Or CRs are trapped between molecular clouds, which have much stronger magnetic fields than the interstellar media and thus form perfect ways to trap CRs just like the CRs are trapped in the field of the earth and form the van Allen radiation belts. Note that  trapping can increase locally the CR density significantly. The larger average distance between MCs allows to trap particles easily up to TeV energies. Trapping improves the isotropy of CRs,  because  the surrounding traps scatter the CRs in all directions, as discussed in detail by \cite{chandran}. Such a trapping picture combined with convection can reproduce the secondary production, the small radial gradient in the production of gamma rays (because the CRs near the source are driven away by the Galactic winds from the disk, thus reducing the strong gamma ray production towards the Galactic center) \cite{breitschwerdt_gammas} and the evidence for Galactic winds   from the ROSAT satellite \cite{rosat}. Additional indirect evidence for the trapping of CRs
comes from the INTEGRAL 511 keV data, as will be discussed in the next section.

The fluxes from CRs from DMA are strongly reduced by even  moderate convection speeds, simply because the CRs drift away from the disk, so any CR in the halo has only a small probability to return to the disk. The difference between the antiproton yield from DMA in isotropic models with its small amount of allowed convection and anisotropic models with convection given by the ROSAT data are compared in Fig. \ref{f3}.
It should be noted that the preliminary antiproton data from PAMELA  are much more precise \cite{pamela}, but consistent with the plotted data from  CAPRICE \cite{caprice} and BESS \cite{bess}.
 In addition, plots concerning the $B/C$ and the $^{10}Be/^9 Be$ ratios are shown. As can be seen the secondary production and the transport time between source and our local solar system can be described in both models. This transport time has to be of the order of $10^7$ years in order to see a significant decrease of the $Be$ ratio, because the unstable $^{10} Be $ has a half life time of $1.6\cdot 10^6$ years, while $^9 Be$ is stable.
The $Be$ ratio increases towards higher energies because the lifetime is proportional to the relativistic $\gamma$-factor, so at high energies less $^{10} Be$ nuclei decay.

\section{The INTEGRAL 511 keV Line}
\label{integral}

The SPI spectrometer on board the INTEGRAL satellite has measured  precisely the line shape and the spatial distribution of the positron-electron annihilation yielding a 511 keV photon line \cite{integral}.
Since positron
annihilation is only efficient at non-relativistic energies, the positrons must
have energies in the MeV range. Sources of such positrons are largely coming from
the decay of radioactive nuclei expelled by dying stars, especially SNIa, since in
this case the core makes up a large fraction of the mass.
This makes it easier for
the positrons, which arise mainly from the $^{56}Co$ decay in the core, to escape from the relatively thin layer of the ejecta, although also here the escape fraction is only a few \% \cite{prantzos}.

Two surprising observations came out of the INTEGRAL data. First of all the width of the line indicated that practically all annihilations take place with electrons from the warm or hot neutral hydrogen gas, not
with electrons from molecular hydrogen, although molecular hydrogen has an averaged density corresponding to more than half of the gas \cite{jean}. In principle one can argue that the filling factor is too low for MCs to be found by positrons. However,  the high magnetic field in the MCs seems to correlate with the interstellar magnetic field \cite{han}, in which case the positrons can spiral towards the MCs. If they do not enter the MCs they are likely to be reflected by the strong gradient of the magnetic field near the MCs.

The second surprise is the high intensity from the region of the Galactic bulge corresponding to an injection rate of approximately $1.5 \cdot 10^{42}~e^+ /s$  in the inner Galaxy~\cite{integral} and a rather low signal from the disk, although the opposite is expected, see e.g. Ref. \cite{prantzos}.
So what happened to the positrons from SNIa explosions in the
disk? In an anisotropic propagation model with convection these positrons are simply blown to the halo by the Galactic winds and here they hardly find an electron to annihilate. Note that at the spot where these positrons are created, the Galactic winds are strong by the pressure created by the SN itself.
The propagation model described above yields indeed that only a small fraction of the positrons stays
inside the disk, thus preventing a strong annihilation signal from the disk.
In the bulge, simply because of its size
of 1-2 kpc, the positrons have enough time to annihilate before reaching the halo.

So the INTEGRAL 511 keV data gives strong constraints on any propagation model, which can be implemented by: i) fast propagation perpendicular to the disk by turbulent diffusion and convection in agreement with ROSAT X-ray data, thus explaining the absence
of a strong annihilation signal in the disk and ii) slow diffusion in the disk, which could happen by trapping of CRs between MCs, thus explaining simultaneously the absence of annihilation in MCs and the long transport time between the sources and our local cavity.
This leads to anisotropic propagation, which is clearly more attractive than isotropic propagation, in which case
the positrons annihilate close to their source \cite{jean1} and one must resort to
new positron sources specific for the bulge, like DMA of very light WIMPS, a possibility which is excluded in most models and requires new physics, like
a $Z^\prime$ boson in order to have fast enough annihilation \cite{boehm}.

\section{The WMAP-haze}\label{wmap}
\begin{figure}
\includegraphics[width=0.45\textwidth,height=0.4\textwidth,angle=0]{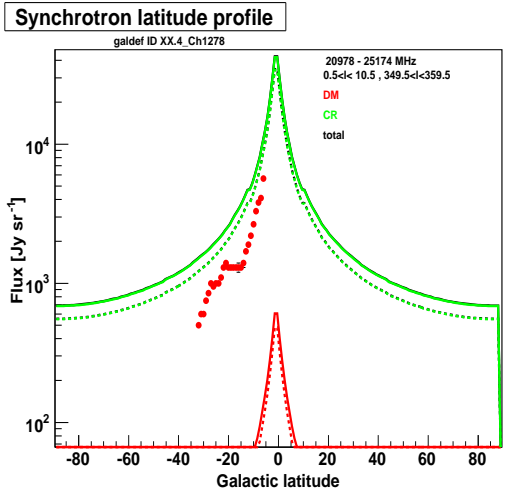}
\caption{The latitude distribution of the synchrotron radiation at 22 GHz in the small longitude range, where the haze has been measured. The top solid curve is the total flux, the red dots represent the haze and the lower curve is the contribution from DMA assuming the same boost factor and same halo as for the gamma ray excess from EGRET.}
\label{f4}
\end{figure}
The WMAP experiment has revealed an excess of microwave emission from the region
around the center of our Galaxy. It has been suggested that this signal, known as
the WMAP-haze, could be synchrotron emission from relativistic electrons and
positrons \cite{haze}. Since the excess is observed towards  the bulge and not in the disk, people
have been inventing sources specific to the bulge and absent in the disk, like
positrons and electrons from  dark matter annihilation in a cuspy halo \citep{haze1}.
However, also the synchrotron radiation from
the disk shows a  steep increase, if one moves from an angle perpendicular to the disk to smaller angles. The different latitude profiles of the total WMAP synchrotron radiation, the haze and the expected contribution from DMA for a cored profile compatible with the EGRET excess are compared in Fig. \ref{f4}. Clearly, a cored profile is as steep as the haze, for which the compatibility with an NFW profile was claimed as support for a signal of DMA. However, the intensity of synchrotron radiation from DMA, compatible with the EGRET excess, is much lower.

Therefore, the original idea that, if there is a  haze it is due to free-free emission from electrons in warm gases seems more likely, especially since the arguments \cite{haze1} of the absence of
 $H_\alpha$ lines, expected in cool gases $(<< 10^4K)$,  the absence
of X-rays from hot gases $T>>10^6$ K and  the absence of gas at the intermediate temperatures (with $T\approx 10^5 K $, because this  is thermally unstable, is not waterproof:  simulations show that gas is usually not in pressure equilibrium and the absorption line of $O-VI$, which traces gas of $T\approx 3\cdot 10^5K$  indicates that this warm gas exists in filamentary structures in agreement with  simulations of a turbulent interstellar medium \cite{breitschwerdt_turbulence}.
The filamentary structure of this warm gas could boost the synchrotron radiation of the free-free emission because of the high local densities.

Summarizing, the priors of an NFW DMA profile  and a warm gas not being present because it is thermally unstable seem  being challenged by better simulations of both, the DM halos and the interstellar medium, so free-free emission can well be present.

\section{Positron fraction}
\begin{figure}
\includegraphics[width=0.45\textwidth,height=0.4\textwidth,angle=0]{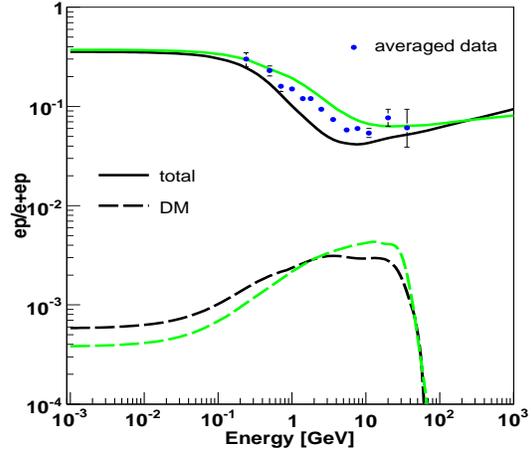}
\caption{The positron fraction measured by HEAT and AMS-01 as  function of energy (averaged data from Ref.~\cite{olzem}).
 The different solid curves correspond to different combinations of electron injection spectra  and  magnetic fields, which are correlated by the fit to the synchrotron radiation spectra of the HASLAM maps and WMAP data. The dashed curves are the corresponding DMA contributions consistent with the EGRET excess
 assuming an equal boost factor for gammas and positrons. The increase in the positron fraction is not due to DMA, but simply due to the different origins: positrons originate from inelastic nucleon scattering, while electrons originate from SN explosions. Therefore the propagation dependence does not cancel, especially in anisotropic models where the electrons have to travel through the disk, thus suffering higher synchrotron radiation losses at higher energies, which leads to an increase in the positron fraction. The preliminary PAMELA data shows a similar increase \cite{pamela}.}
\label{f5}
\end{figure}
The positrons from DMA are produced mainly by the decays of positively charged pions produced after the hadronization of the quarks. The background comes from positive pions produced by inelastic collisions of CRs with the gas in the disk. Electrons are also produced by the decays of the negative pions produced in the same processes, but this is only a small fraction compared to the electrons from SNRs. Therefore the electrons originate from a quite different source and position in the Galaxy than the positrons. Therefore the propagation model dependence does {\it not} cancel in the positron fraction, defined as $e^+/(e^+ + e^-)$. The standard GALPROP with isotropic propagation predicts a falling curve for this ratio, while the data increases above 7 GeV, as shown in Fig. \ref{f5}. An increase in the positron fraction can be  obtained as well after including  anisotropic
propagation in GALPROP, as shown by the curves in Fig. \ref{f5}. The physics is clear: 
electrons in an anisotropic model diffuse mainly from the sources through the disk, since if they diffuse into the halo they are carried away by convection. These electrons suffer thus more synchrotron losses
than positrons, since the latter are produced locally to a larger extent. In  isotropic propagation models the electrons and positrons are mostly collected from the halo, so here the synchrotron losses play less of a role. The energy loss curves of electrons peak at low energies from ionization losses and Bremsstrahlung and at high energies from synchrotron radiation.  This depletes the electron spectra at low and high energies and the corresponding minimum in the positron fraction depends  on the propagation parameters and the averaged magnetic field.

The contribution from DMA is small given the observed excess of Galactic gamma rays and taking the same boost factor  also for the positrons, as shown by the lower curves in Fig. \ref{f5}. This is expected, since most of the positrons from DMA are produced in the halo, thus drifting away by convection, while the background is produced mainly in the disk.
For antiprotons the background is reduced by the large threshold of CR protons, which requires protons  to be effectively above 10 GeV, while for the light pions there is hardly any threshold effect. Therefore the relative DMA contribution   is expected to be  larger for antiprotons than for positrons. Most propagation models predict  too few antiprotons \cite{strong_rev}, if one does not resort to so-called optimized models, which are tuned to reproduce the EGRET excess of gamma rays by increasing the density of high energy protons and electrons. The higher density of energetic protons increases also the antiproton flux. The difficulty with this model is, however, that the EGRET data below 1 GeV are well described by the locally observed CR spectra, so one has to introduce an ad hoc large break in the injection spectra to increase the electron and proton CR density at high energies outside our local cavity without changing the low energy part. Such a break in the relativistic regime is not expected from the acceleration of protons by SN explosions \cite{berezinsky}. Given that the energy loss time of CR protons is much longer than the residence time in our Galaxy one expects the same shape of the CR proton spectra everywhere, as born out by the fact that the EGRET data can be well fitted  with the locally observed CR shape  in all sky directions \cite{us}.
In addition, the high energy electrons and high energy protons have to be enhanced by different factors
(4 and 2, respectively \cite{om}), which is hard to explain by propagation models, since these would enhance CR densities for high energy electrons and protons in a similar way.

\section{Conclusion}\label{conclusion}
The various DMA signals for charged particles and gamma rays have been discussed. The existing propagation models assume the same propagation in the halo and the disk, which turn out to be incompatible with the evidence for convection from the ROSAT data, the INTEGRAL data on the large bulge/disk ratio for positron annihilation and the small radial gradient of diffuse gamma rays, as observed by EGRET. However, if one allows for  faster propagation in the halo than in the disk all these new observations can be brought inline with all other observation concerning secondary production (B/C), cosmic clocks $(^{10}Be/^9Be)$
and synchrotron radiation. The only exceptions are the too low antiproton flux and the EGRET excess of diffuse gamma rays above 1 GeV, but these can be beautifully remedied by the annihilation signal of a WIMP in the range of 50-100 GeV without resorting to ad hoc optimized models. The  positron fraction, the WMAP haze and the large bulge/disk ratio of low energy positron annihilation are all explained well without DMA in the anisotropic propagation model.  Finally it should be emphasized that the preliminary positron and antiproton data from PAMELA are in perfect agreement with the EGRET excess interpreted as DMA and the anisotropic propagation required by the ROSAT data. Especially the rising positron fraction is easily explained in anisotropic propagation models without DMA.

\section{Acknowledgments}
 I wish to thank my close collaborators Iris Gebauer, Markus Weber, Dmitri Kazakov and Valery Zhukov for helpful
discussions.
 This work was supported by the BMBF (Bundesministerium f\"ur Bildung und Forschung) via the DLR
(Deutsches Zentrum f\"ur Luft- und Raumfahrt).

\end{document}